\documentclass[twocolumn,showpacs,preprintnumbers,amsmath,amssymb]{revtex4}

\usepackage{graphicx} 
\usepackage{dcolumn} 
\usepackage{bm} 

\begin{document}


\title{Double non-equivalent chain structure on vicinal Si(557)-Au surface}

\author{M. Krawiec}

 \email{krawiec@kft.umcs.lublin.pl}

\author{T. Kwapi\'{n}ski}
\author{M. Ja\l ochowski}

\affiliation{Institute of Physics and Nanotechnology Center, 
             M. Curie-Sk\l odowska University, Pl. M. Curie-Sk\l odowskiej 1,
	     20-031 Lublin, Poland}

\date{\today}

\begin{abstract}
We study electronic and topographic properties of the vicinal Si(557)-Au 
surface using scanning tunneling microscopy and reflection of high energy 
electron diffraction technique. STM data reveal double wire structures along 
terraces. Moreover behavior of the voltage dependent STM tip - surface distance 
is different in different chains. While the one chain shows oscillations of the 
distance which are sensitive to the sign of the voltage bias, the oscillations 
in the other chain remain unchanged with respect to the positive/negative 
biases. This suggests that one wire has metallic character while the other 
one - semiconducting. The experimental results are supplemented by theoretical 
calculations within tight binding model suggesting that the observed chains are 
made of different materials, one is gold and the other one is silicon chain.
\end{abstract}

\maketitle


\section{\label{introduction} Introduction}

Studies of the high-index (vicinal) surfaces have attracted much attention as
templates for formation of the low dimensional structures on them 
\cite{Himpsel}. Such structures, usually one-dimensional, are important from
scientific point of view as they allow to study phenomena like Luttinger liquid 
\cite{Luttinger,Haldane}, Peierls instability \cite{Peierls}, self-assembling 
of arrays of nanowires \cite{Jalochowski} or basic properties of electrons. On 
the other hand they have potential technological applications in 
nanoelectronics or quantum computing.

There are many examples of the vicinal surfaces studied recently, including
Au decorated Si(335) surface \cite{Zdyb,Crain,Crain_1,MK_1}, clean Si(557) 
\cite{Henzler}, Si(557)-Au \cite{Crain_1,Jalochowski_1,Segovia,Losio,Altmann,
Sanchez,Robinson,Sanchez_1,Okino}, Si(557)-Ag \cite{Zhachuk} or Si(5512)-Au 
\cite{Baski,Lee,Ahn,Jeong}. All these surfaces allow to grow on them one 
dimensional wires along terraces. 

In the case of the Si(335)-Au surface, which consists of Si(111) terraces 
$3\frac{2}{3} \times a_{[1 1 \bar{2}]}$ wide, a single long mono-atomic chain on
each terrace has been observed at gold coverage $0.28$ ML. This surface has 
been investigated using reflection of high energy electron diffraction (RHEED) 
\cite{Zdyb,MK_1}, angle resolved photoemission spectroscopy (ARPES) 
\cite{Crain,Crain_1} and scanning tunneling microscopy (STM) 
\cite{Crain_1,MK_1} indicating metallic character of the wires grown on it.
Moreover, it was found experimentally and confirmed theoretically that these 
wires show bias dependent oscillations of the STM tip - surface distance along 
wire with a period of $2 \times a_{[1 \bar{1} 0]}$ \cite{MK_1}.

Ja\l ochowski {\it et al.} \cite{Jalochowski_1} showed how small amount of gold 
stabilizes Si(557)-Au surface. This surface possesses wider terraces 
($5\frac{2}{3} \times a_{[1 1 \bar{2}]}$) and at gold coverage $0.2$ ML one
observes not single but double mono-atomic chains on each terrace 
\cite{Jalochowski_1,Segovia,Losio,Altmann,Sanchez,Robinson,Sanchez_1,Okino}. 
Many experimental techniques have been applied to study properties of this 
surface. Segovia {\it et al.} \cite{Segovia} using ARPES claimed that the wires 
show Luttinger liquid (LL) behavior which turned out later on to be not LL but 
usual Fermi liquid with two metallic bands \cite{Losio,Altmann}. Those 
conclusions have been reached by studying it using ARPES and STM techniques and 
confirmed theoretically within density functional theory \cite{Sanchez}. On the 
other hand, conductivity measurements suggest non-metallic character of the 
surface \cite{Okino}, which seems to be in contradiction with ARPES data. 
However, one has to have in mind that the chains are not ideal but they are 
discontinued (tens of nanometers long) so even along wire there are metallic as 
well as semiconducting regions, which may explain discrepancy between 
conductivity measurement and ARPES data. Moreover, it turned out that such 
amount of gold is sufficient to form single wire only \cite{Sanchez_1,Crain}, 
unless we assume that gold atoms are distributed with distance equal to twice 
of the lattice constant in direction $[1 \bar{1} 0]$. Thus it is enough of gold 
to form two chains. However the existence of the single Au chain per terrace 
has been found using x-ray diffraction \cite{Robinson}. 

In the present work we use scanning tunneling microscopy (STM) technique to 
study gold induced structures on Si(557) surface. Those experimental studies 
are supplemented by theoretical ones based on tight binding model where the STM 
tunneling current is calculated using non-equilibrium Keldysh Green function 
formalism. Theoretical description remains in good qualitative agreement with 
the STM data. Rest of the paper is organized as follows: in Sec. 
\ref{experimental} we describe the experimental setup and provide some data. In 
Sec. \ref{model} we introduce model of the chains on surface and the results 
are presented in Sec. \ref{results}. The comparison of the theoretical 
calculations with the experimental data is given in Sec. \ref{comparison}. 
Finally Sec. \ref{conclusions} contains some conclusions.


\section{\label{experimental} Experimental}

Experimental setup consists of ultra high vacuum (UHV) chamber equipped with a
scanning tunneling microscope (type OmicronVT) and reflection high energy
electron diffraction (RHEED) apparatus. Samples were prepared in-situ and the
base pressure was less than $5 \times 10^{-11}$ mbar during measurements. The
one-dimensional structures have grown after deposition of $0.2$ ML of Au, 
heating the sample at temperature $950$ K for $20$ s and gradually annealing 
to the room temperature for $3$ min. The quality of the surface reconstruction 
has been controlled by RHEED technique. All STM data have been collected at 
room temperature.

\begin{figure}[h]
 \resizebox{0.8\linewidth}{!}{
  \includegraphics{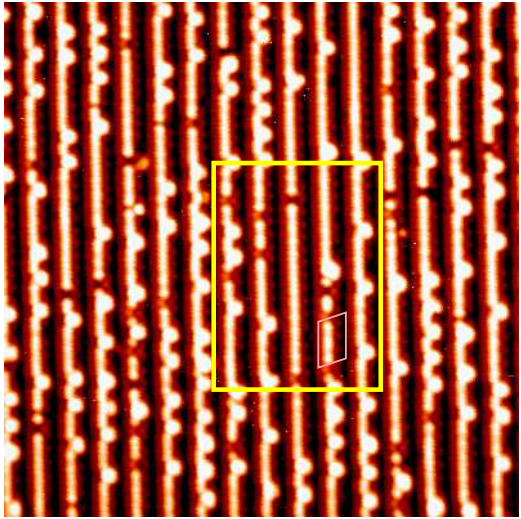}
  }
 \caption{\label{Fig1} (Color online) The $30 \times 30 \; nm^2$ area STM 
          image of the Si(557) - $0.2$ ML Au surface recorded with tunneling 
	  current $I = 0.05$ nA and sample bias equal to $U = - 1.0$ V. The 
	  frames mark areas chosen for further analysis.}
\end{figure}

Figure \ref{Fig1} shows large area STM image of the Si(557) surface covered 
with $0.2$ ML of gold. 

\begin{figure}[h]
 \resizebox{0.9\linewidth}{!}{
  \includegraphics{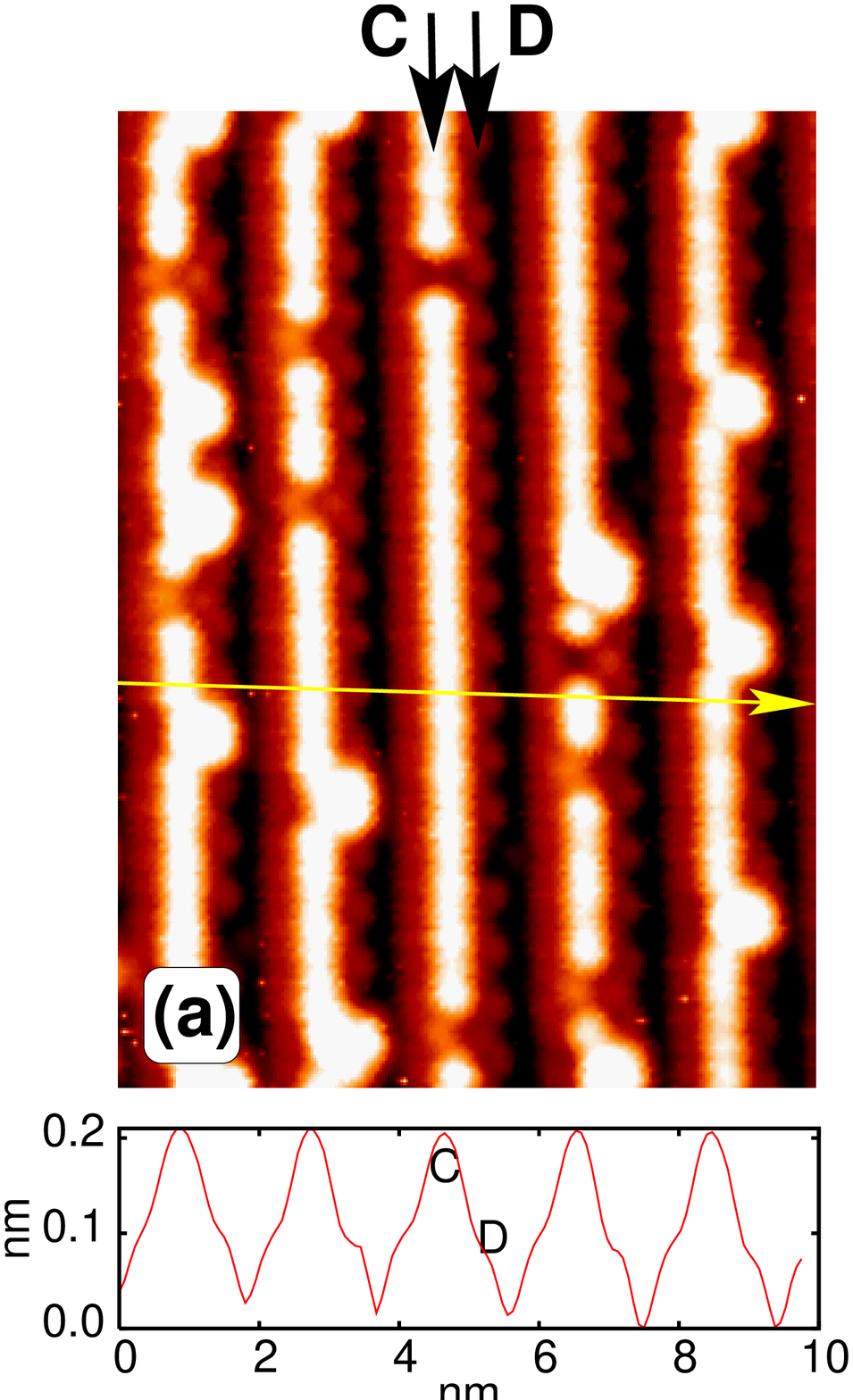}
  \hspace{0.2\linewidth} \includegraphics{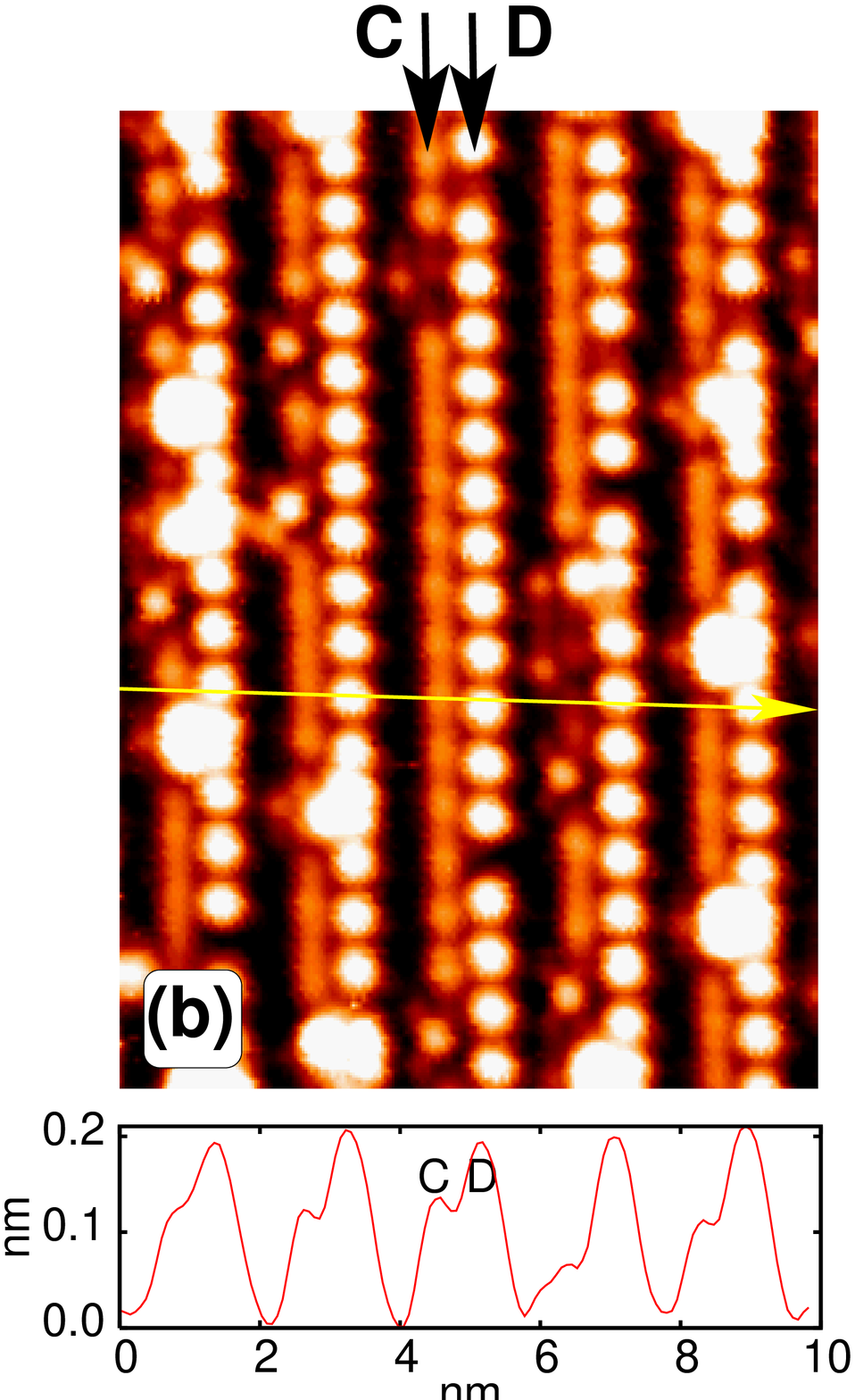}
  }
 \caption{\label{Fig2} (Color online) The $9.75 \times 13.5 \; nm^2$ STM 
          topography images of the same area of the Si(557)-Au surface recorded 
	  at two different sample biases $U = -1.0 \; V$ (a) and $U = 1.0 \; V$ 
	  (b) with the tunneling current $I = 0.05 \; nA$. Bottom panels show 
	  profile lines perpendicular to the chains, indicated by long arrows 
	  in the main panels.}
\end{figure}

Figure \ref{Fig2} shows enlarged part of the image marked with thick line in 
Fig. \ref{Fig1} recorded at tunneling current $I = 0.05$ nA and the sample bias 
$U = - 1.0$ V (a) and $+ 1.0$ V (b).
We note that images (a) and (b) represent the same area of the sample. The STM
images clearly show sets of double chains on each terrace. Distance between
these sets is equal to the width of the unreconstructed Si(557) surface 
terrace, $5\frac{2}{3} \times a_{[1 1 \bar{2}]}$ ($1.92$ nm), and the distance 
between two chains on the same terrace is equal to $2 \times a_{[1 1 \bar{2}]}$
($0.68$ nm). However the chains on the same terrace show different behavior. 
While the chain indicated by arrows and denoted as C in Fig.\ref{Fig2} (a) and 
(b) is apparently continuous and very similar to that grown on Si(335)-Au 
surface \cite{MK_1}, the wire D differs considerably and reminds a row of well 
ordered single atoms rather than one-dimensional object. However periodicity 
along chains are the same in both wires and equal to 
$2 \times a_{[1 \bar{1} 0]}$ ($0.77$ nm). It is worth to note that similar
periodicity along the chain has been also found in different structures, like
Si(335)-Au \cite{MK_1} with single chain per terrace or 
Si(111)-(5 $\times$ 2)-Au \cite{Erwin} with double Au chains. 

\begin{figure}[h]
 \resizebox{0.8\linewidth}{!}{
  \includegraphics{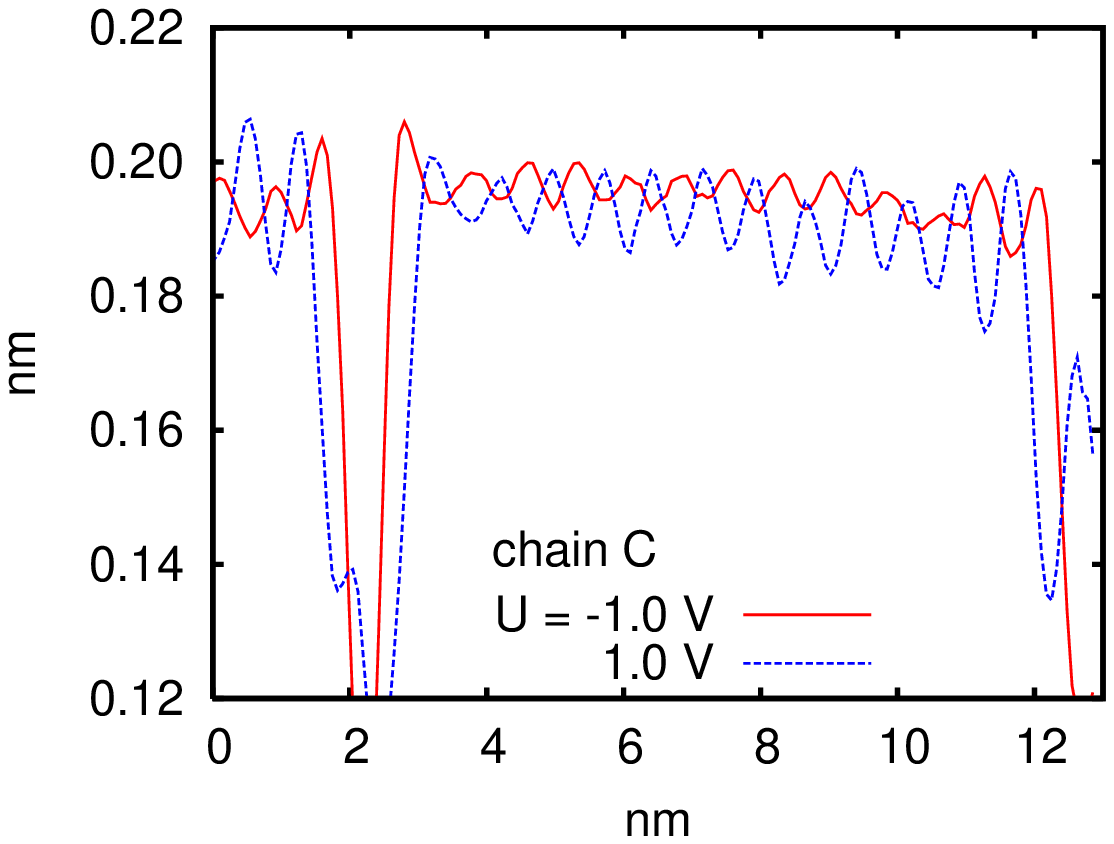}
  }
 \resizebox{0.8\linewidth}{!}{
  \includegraphics{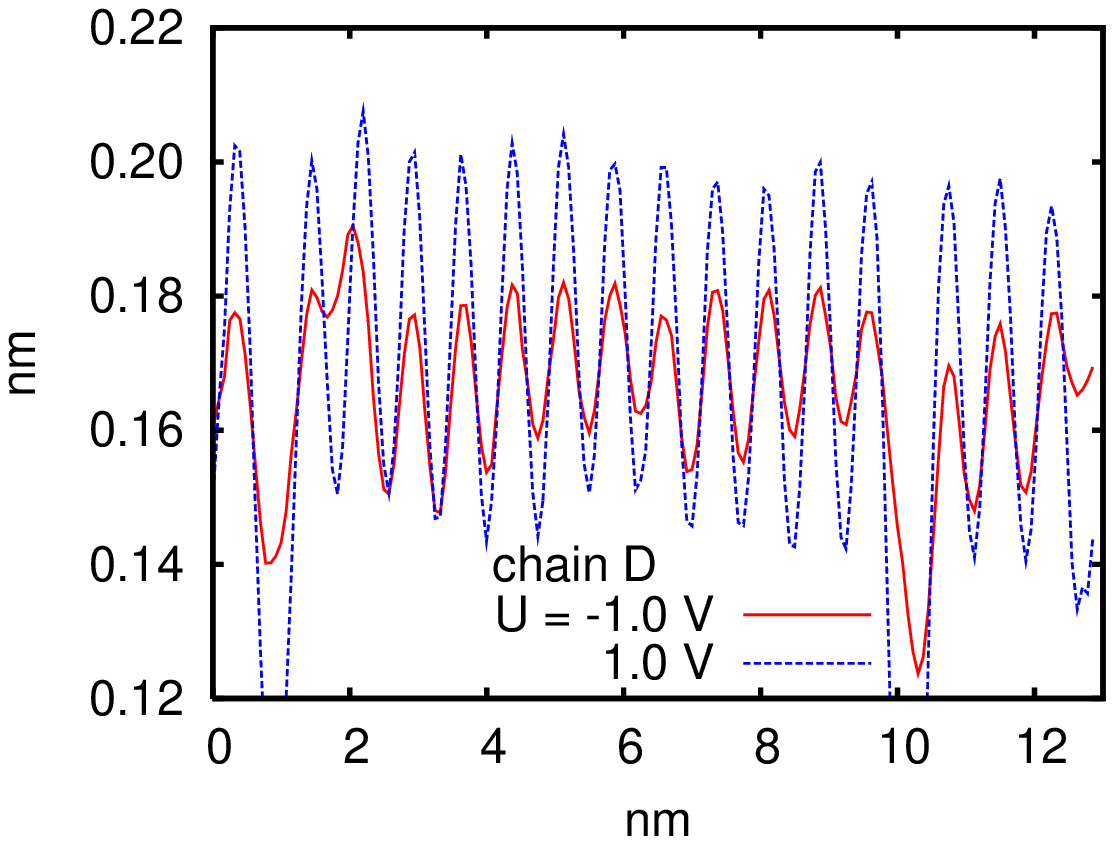}
  }
 \caption{\label{Fig3} (Color online) Cross sections along middle chains 
          indicated by short arrows in Fig. \ref{Fig2} (a) and (b). Top panel 
	  shows cross section along chain C for bias voltage $U = - 1.0$ V 
	  (solid line) and $+ 1.0$ V (dashed line), the bottom panel - along 
	  chain D.}
\end{figure}

Figure \ref{Fig3} displays cross sections along atomic chain C (a) and D (b) 
indicated by arrows in Fig.\ref{Fig2}. 
As one can read from the figure both chains show oscillations of the electron 
density with distance along their length. Moreover, it is clearly seen that 
those chains behave differently upon reversing the sample bias. Topography of 
the nanowire C (top picture) reverses (maxima change into minima and 
vice versa) with switching of the voltage bias from negative to positive 
values. Such behavior has been previously observed in the case of Si(335)-Au 
structure \cite{MK_1} and has strictly electronic origin. On the other hand 
there is a lack of such reversing in the chain D where its topography remains 
almost unchanged, except that at positive bias the structures are higher. We 
stress that in order to exclude possible influence of the sample thermal drift 
the presented STM images were recorded simultaneously, with negatively biased 
sample during line forward scan and positively biased during line backward 
scan. 

\begin{figure}[h]
 \resizebox{\linewidth}{!}{
  \includegraphics{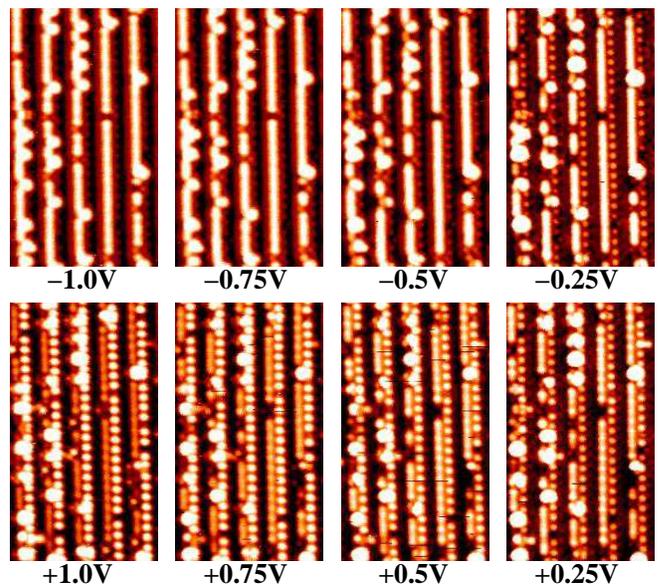}
  }
 \caption{\label{Fig4} (Color online) The $9.6 \times 19.1$ nm$^2$ STM 
          topography images of the same area of the Si(557)-Au surface recorded 
	  at different sample biases (indicated below the images) with the 
	  tunneling current $I = 0.05 \; nA$.}
\end{figure}

It is interesting to see the changes of topography with the bias voltage. The 
evolution of topography changes with the bias voltage is shown in Fig. 
\ref{Fig4}. 
We observe that the chains of C (D) type simultaneously become less (more) 
visible while going from negative to positive biases. Moreover, chains of D 
type do not show any topography reverse. On the other hand, there is a reverse 
of the topography of the C type chains when crossing zero bias. 

Another interesting effect is associated with recently observed zero 
dimensional end states \cite{Crain_2}. Such states must have the wave function
localized to the end atoms in a chain, and it must must decay exponentially 
into the chain \cite{Davison,Crain_2}. Similarly for Si(553)-Au surface
\cite{Crain_2}, we observe that C type chains at negative sample bias appear 
longer than at positive bias (see Fig. \ref{Fig3} - top panel). At negative 
bias the end atoms are well visible while for negative bias the atoms second 
form the end are enhanced and the end atoms became supressed. This fact can be 
explained in terms of different densities of states above and below the Fermi 
energy (see discussion in Sec. \ref{results}). However, unlike in Ref. 
\cite{Crain_2}, the topography modulation along chain is twice a lattice 
constant ($2 \times a_{[1 \bar{1} 0]} = 0.77$ nm) and what is important we also 
observe the topography reverse. We stress that there is no such an effect for 
chains of D type (see Fig. \ref{Fig3} - bottom panel). 

\begin{figure}[h]
 \resizebox{0.8\linewidth}{!}{
  \includegraphics{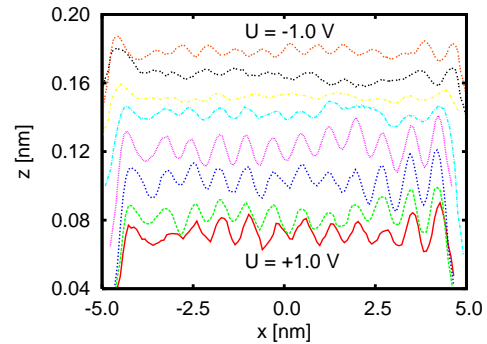}
  }
 \caption{\label{Fig5} (Color online) Cross sections along the longest chain 
          indictated by arrow with label C in Fig. \ref{Fig2} for different 
	  bias voltages: $U = -1.0$, $-0.75$, $-0.5$, $-0.25$, $0.25$, $0.5$,
	  $0.75$, $1.0$ from top to bottom. The curves are shifted for better
	  presentation.}
\end{figure}

To see the evolution of the topography of chain C type with the sample bias we 
show in Fig. \ref{Fig5} the cross sections along the longest chain indicated 
by arrow with label C in Fig. \ref{Fig2}.
It is clearly visible how the length of the chain becomes shorter going from
negative to positive voltages. Moreover around zero bias we observe the reverse
of the topography. 

\begin{figure}[h]
 \resizebox{0.8\linewidth}{!}{
  \includegraphics{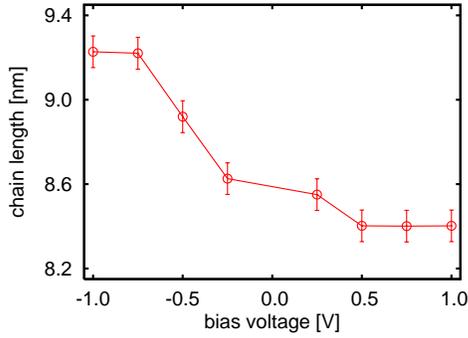}
  }
 \caption{\label{Fig6} (Color online) The length (distance between endmost
          maxima in Fig. \ref{Fig3}) of the longest chain C indicated by arrow 
	  in Fig. \ref{Fig2} as a function of the bias voltage. }
\end{figure}

Figure \ref{Fig6} shows the length of above discussed the chain as a function of
the sample bias. 
We define the length as a distance between endmost maxima in the topography 
profile (see Fig. \ref{Fig5}). We observe that the effect associated with the 
length changing of the chain takes place in the range between $-0.75$ V to 
$+0.5$ V. Below the bias $-0.75$ V and above $+0.5$ V it remains almost 
constant, and the length difference between those biases is equal $0.77$ nm 
(twice a lattice constant along the chain).

From the experimental data we cannot judge what material (Si or Au) those wires 
are composed of. Certainly there is no sufficient amount of gold to produce two 
gold chains within terrace, unless we assume that gold atoms occupy every 
second site in direction $a_{[1 \bar{1} 0]}$, which would be supported by fact 
that those chains show oscillations with period twice as large as the lattice 
constant in this direction. However it is difficult to find any particular 
reason supporting this assumption. So one can assume that we deal with one gold 
and one silicon chains. This would be supported by above-mentioned differences 
between chains seen in STM images. However there might be another scenario in 
which both structures come from silicon whilst gold atoms substitute into the 
surface producing a third chain not visible in STM images. Such supposition is 
supported by the fact that Au is more electronegative than Si and binds one Si 
electron with their $s$, $p$ electrons in a low laying state \cite{Sanchez}, 
according to the first principle calculations \cite{Crain_1,Sanchez_1}. The 
chain structures visible in STM are likely to be Si atoms at step edge 
(chain C), similar to those in Si(335)-Au, and Si adatoms (chain D).


\section{\label{model} Theoretical description}

The experimental studies suggest that the electronic states in the chain C 
have extended nature while in chain D they are localized. In the language of 
the tight binding model the hopping integral between atoms in chain C has 
larger value than that in the other chain or single particle electron energies 
of the chain C atoms lay much closer to the Fermi energy of the surface than 
those of the chain D. So one can conclude that chain C has metallic character 
whilst D is of isolating (semiconducting) nature. Moreover, the fact that at 
negative sample bias $U$ (Fig. \ref{Fig2} a)) chain C is more visible in the 
STM topography image than the chain D while at positive $U$ (Fig. 
\ref{Fig2} b)) the effect is opposite, suggests that we deal with mainly 
occupied states in the chain C and empty states in D. 

\begin{figure}[h]
 \resizebox{0.8\linewidth}{!}{
  \includegraphics{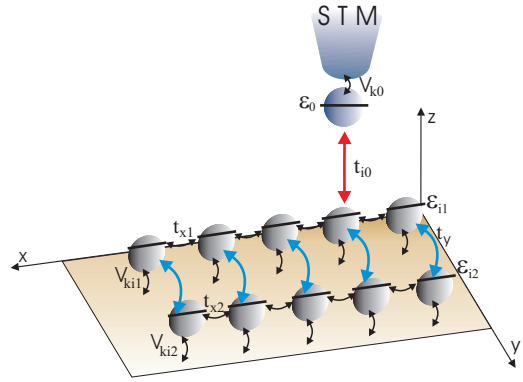}
  }
 \caption{\label{Fig7} (Color online) Schematic view of the model STM system 
          containing two chains $\alpha =1$, $2$, composed of atoms with single 
	  particle energies $\varepsilon_{i\alpha}$, intra chain hopping 
	  $t_{{\text{x}} \alpha}$ and inter chain hopping $t_{\text{y}}$ 
	  coupled to the surface via $V_{{\bf k} i \alpha}$. STM tip is modeled 
	  by single atom with energy $\varepsilon_0$ coupled to the STM 
	  electrode via $V_{{\bf k} 0}$. Parameter $t_{i 0}$ is responsible for 
	  tunneling between STM tip and the chains.}
	   
\end{figure}

In order to explain the effect associated with the electron density 
oscillations and differences between neighboring chains observed in STM 
images, we propose a model of two coupled non-equivalent chains composed of Au
(chain C) and Si (chain D), schematically depicted in Fig. \ref{Fig7}. 

The Hamiltonian describing system composed of surface, two chains and STM tip 
is:
\begin{eqnarray}
H = H_{\text{STM}} + H_{\text{tip}} + H_{\text{wire}} + H_{\text{surf}} + 
H_{\text{int}}
\label{Hamiltonian}
\end{eqnarray}
where 
\begin{eqnarray}
H_{\text{STM}} = \sum_{{\bf k} \in \text{STM}} \epsilon_{\bf k} c^+_{\bf k}
c_{\bf k}
\label{H_STM}
\end{eqnarray}
is for STM electrode electrons with single particle energies 
$\epsilon_{\bf k}$. The STM tip is modeled by single atom with energy level 
$\varepsilon_0$ 
\begin{eqnarray}
H_{\text{tip}} = \varepsilon_0 c^+_0 c_0
\label{H_d}
\end{eqnarray}
The wire part of the Hamiltonian is described by
\begin{eqnarray}
H_{\text{wire}} = \sum_{i \alpha} \varepsilon_{i\alpha} c^+_{i\alpha}
c_{i\alpha} + \sum_{{\text{x}}\alpha} t_{{\text{x}}\alpha} 
c^+_{i\alpha} c_{j\alpha} +
\sum_{i\alpha \neq \beta} t_{\text{y}} c^+_{i\alpha} c_{i\beta}
\label{H_wire}
\end{eqnarray}
where $\varepsilon_{i\alpha}$ is the atomic energy of the $i$-th atom in chain 
$\alpha = 1,2$, $t_{{\text{x}}\alpha}$ is the hopping integral between 
neighboring atoms in the same chain, while $t_{\text{y}}$ - hopping between 
neighboring atoms in different chains. Hamiltonian of the surface is
\begin{eqnarray}
H_{\text{surf}} = \sum_{{\bf k} \in \text{surf}} \epsilon_{\bf k} c^+_{\bf k}
c_{\bf k}
\label{H_surf}
\end{eqnarray}
and interactions between different subsystems are in the form
\begin{eqnarray}
H_{\text{int}} = \sum_{{\bf k} \in \text{STM}} (V_{{\bf k}0} c^+_{\bf k} c_0 + 
\text{h.c.}) + \sum_{i \alpha} (t_{i0\alpha} c^+_{i\alpha} c_0 + \text{h.c.}) 
\nonumber \\
+ \sum_{{\bf k} \in \text{surf}} \sum_{i\alpha} (V_{{\bf k} i\alpha} 
c^+_{\bf k} c_{i\alpha} + \text{h.c.})
\label{H_int}
\end{eqnarray}
with $V_{{\bf k}0}$ being hybridization between STM electrode electrons and 
STM tip, $t_{i0\alpha}$ - hopping between atoms in a chain $\alpha$ and STM tip 
and finally $V_{{\bf k} i\alpha}$ - hybridization matrix element connecting 
surface and the chain $\alpha$. In all above equations we have omitted spin 
index in the electron creation (annihilation) operators as we are not 
interested in magnetic properties and assume that system is in paramagnetic 
phase. 

In order to calculate the tunneling current from the STM electrode to the
surface we follow the standard derivations \cite{Haug,Niu,MK_2} and get
\begin{eqnarray}
I = \frac{2e}{\hbar} \int^{\infty}_{-\infty} \frac{d\omega}{2\pi}
{\text{T}}(\omega) [f_{\text{STM}}(\omega) - f_{\text{surf}}(\omega)]
\label{curr_final}
\end{eqnarray}
where $f_{\text{STM (surf)}}(\omega) = 
(\exp{((\omega - \mu_{\text{STM (surf)}})/kT)}+1)^{-1}$ is the Fermi 
distribution function of the STM (surf) electrode with the chemical potential 
$\mu_{\text{STM (surf)}}$ and the transmittance ${\text{T}}(\omega)$ is given 
in the form
\begin{eqnarray}
{\text{T}}(\omega) = \Gamma_{\text{STM}}(\omega) \Gamma_{\text{surf}}(\omega) 
|\sum_{i\alpha} G^r_{i d \alpha}(\omega)|^2
\label{transmit}
\end{eqnarray}
$\Gamma_{\text{STM}}(\omega) = 2\pi \sum_{{\bf k} \in \text{STM}} 
|V_{{\bf k} 0}|^2 \delta{(\omega - \epsilon_{\bf k})}$ and 
$\Gamma_{\text{surf}}(\omega) = 2\pi \sum_{{\bf k} \in \text{surf}} 
|V_{{\bf k} \alpha}|^2 \delta{(\omega - \epsilon_{\bf k})}$ is the coupling 
parameter between STM electrode and the tip atom and surface and the chain 
respectively. $G^r_{i d \alpha}(\omega)$ is the the Fourier transform of the 
retarded GF 
$G^r_{i d \alpha}(t) = i \theta(t) \langle [d_i(t), d^+(0)]_+ \rangle$, and is 
the matrix element (connecting the tip atom $0$ with $i$-th atom in chain 
$\alpha$) of full GF, solution of the equation
\begin{eqnarray} 
(\omega \hat 1 - \hat H) \hat G^r(\omega) = 1
\label{EOM}
\end{eqnarray}
Full GF $\hat G^r(\omega)$ is $(2N +1) \times (2N + 1)$ matrix ($N$ atoms in a
wire $1$, $N$ atoms in a wire $2$ and the tip site), which has to be determined,
just by inverting matrix $(\omega \hat 1 - \hat H^{-1})$. Note that within the 
present model there are no Coulomb interactions between electrons and the
problem can be solved exactly.


\section{\label{results} Numerical results}

So far we are able to calculate the tunneling current for a given bias voltage
$eU = \mu_{\text{STM}} - \mu_{\text{surf}}$ at constant STM tip - surface
distance. In our model, the quantity which is related to this distance is the
coupling between tip atom and the chain $\alpha$ 
$\Gamma_{i0 \alpha} \equiv |t_{i 0 \alpha}|^2$ (in units of 
$\Gamma = \Gamma_{\text{STM}} + \Gamma_{\text{surf}} = 1$). From the physics of 
tunneling phenomena we know that $\Gamma_{i0 \alpha}$ exponentially depends on
distance $z$ between STM tip and the chain. A rough estimate of 
$\Gamma_{i0 \alpha}$ may be deduced by taking both the tip and the chain 
orbitals as the lowest states of square potential wells of spatial widths 
$l_M$ separated by the barrier of height equal to the work function and width 
equal to the distance $z$ between STM tip and the chain. The half of the ground 
state energy splitting in this double well structure $\Gamma_{i0 \alpha}$ is 
thus given by \cite{Calev}: 
\begin{eqnarray}
\Gamma_{i 0 \alpha} = \frac{\hbar^3 \pi^2 \sqrt{2 m W_f}}
{m^2 l^3_M \left( \frac{\hbar^2 \pi^2}{2 m l^2_M} + W_f \right)} 
e^{-\frac{z}{\hbar} \sqrt{2 m W_f}} 
\label{Gamma_z}
\end{eqnarray}
where $m$ is the electron mass, $W_f$ - work function taken as 
$4.5 \Gamma$, $l_M$ - length parameter (of order of an orbital spatial size) 
chosen as $0.52$ in units of the lattice constant $a_{[1 \bar{1} 0]}$. 

In numerical calculations we have chosen the Fermi energy of the surface as 
zero of energy scale ($E_{\text{F}} = 0$). All energies are measured with 
respect to it in units of 
$\Gamma = \Gamma_{\text{STM}} + \Gamma_{\text{surf}} = 1$ and for simplicity we
have chosen $\Gamma_{\text{STM}} = \Gamma_{\text{surf}} = 0.5$. The quantities 
concerning distances are expressed in units of the lattice constant along chain 
($a_{[1 \bar{1} 0]} = 1$). Moreover, the distance between chains is modeled via 
inter-chain hooping integral $t_{\text{y}}$. 

To get the information on topography of the wire we have solved
Eqs.(\ref{curr_final})-(\ref{EOM}) self-consistently for a given bias voltage 
$eU$ and fixed current $I$, got value of $\Gamma_{i0 \alpha}$ and finally 
determined $z$ from relation (\ref{Gamma_z}). 

\begin{figure}[h]
 \resizebox{0.8\linewidth}{!}{
  \includegraphics{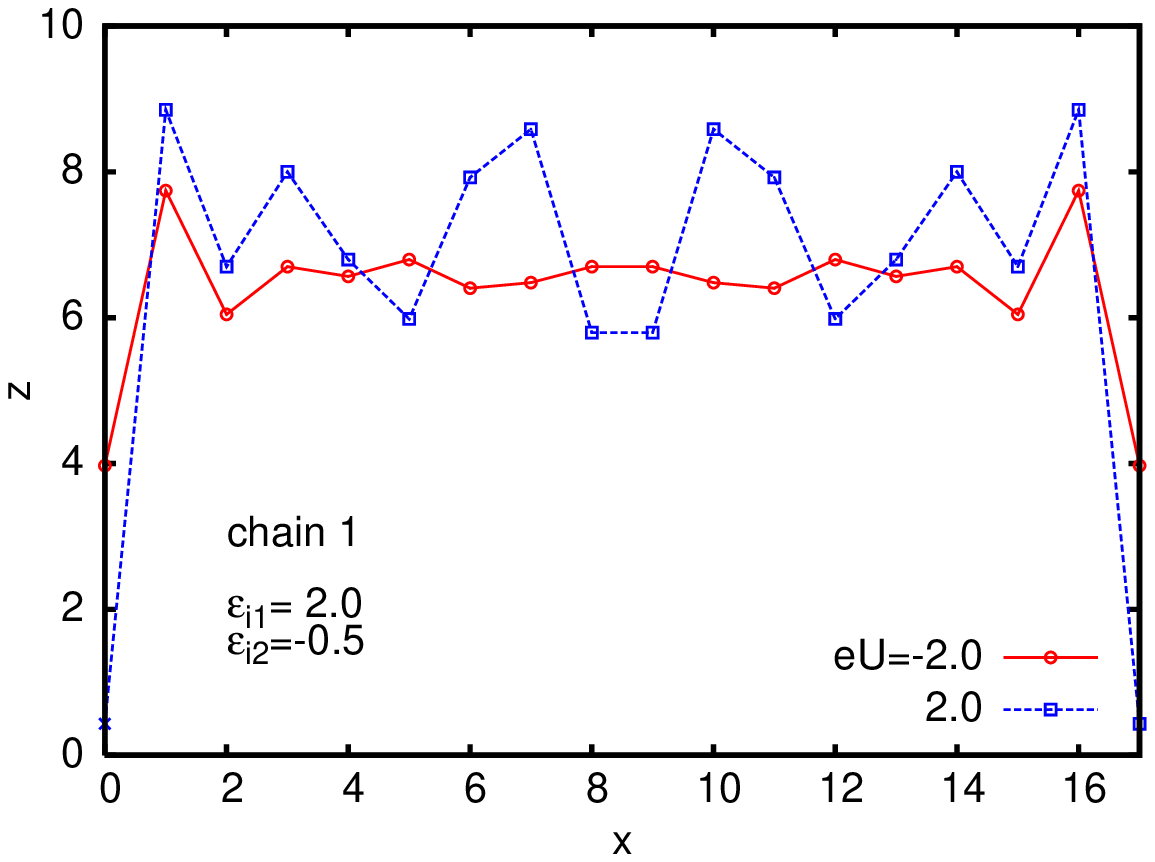}
  }
 \resizebox{0.8\linewidth}{!}{
  \includegraphics{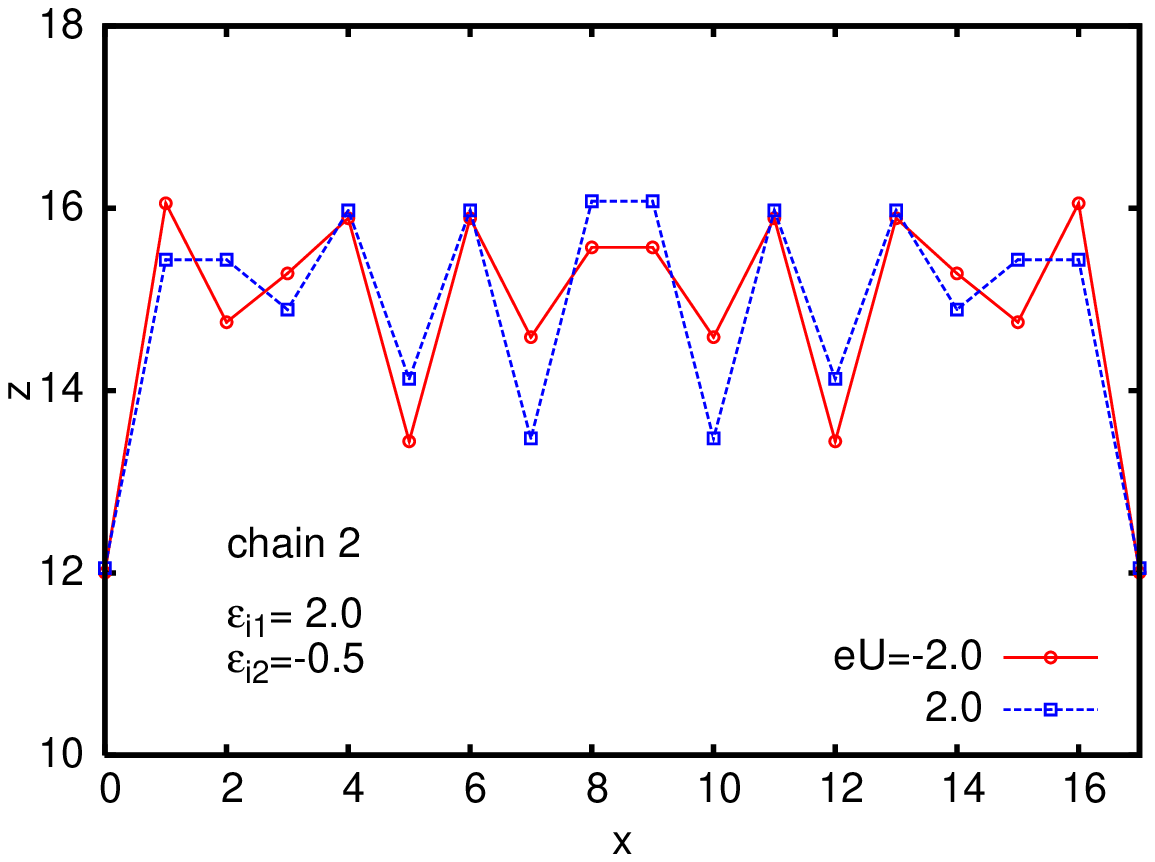}
  }
 \caption{\label{Fig8} (Color online) STM tip - surface distance along chain 
          $1$ (top panel) and chain $2$ (bottom panel) for $eU = -2.0$ 
	  (circles) and $eU = 2.0$ (squares). The other parameters are: 
	  $\varepsilon_{i 1}=2.0$, $\varepsilon_{i 2} =-0.5$, 
	  $\varepsilon_0 = 0.0$, $t_{\text{x}} = 1.0$ and 
	  $t_{\text{y}} = 5 \cdot 10^{-2}$. Note that point represent positions 
	  of the STM tip (exactly above chain atoms) and the line is a guide 
	  for the eye.}
\end{figure}

In Fig. \ref{Fig8} we show the comparison of the STM tip - surface distance 
$z$ along chains $1$ and $2$ for negative and positive sample biases. 
It is clearly seen that chain $1$ (corresponding to the chain C observed in
experiment) shows the bias dependent reverse of the topography (top panel) 
while chain $2$ (chain D) is insensitive to it (bottom panel). Furthermore, if 
the atomic energies of the chain are close to the Fermi energy, the topography 
of the chain remains almost unchanged with respect to the reverse of the bias 
voltage, while it reverses if the chain atomic energies lay far away from the 
Fermi energy. We have performed a number of calculations for various values of 
$\varepsilon_{i \alpha}$ and this seems to be a general tendency, provided the 
other parameters are the same for both chains. Actually this can be understood 
from a simple argument if we consider not a chain but single atom with atomic 
energy $\varepsilon_{\text{at}}$. Namely, the atomic local density of states 
(LDOS) has Lorentzian shape due to the coupling to the STM tip and the surface, 
and if $\varepsilon_{\text{at}}$ is close to the Fermi energy, the LDOS is 
almost symmetric with respect to the Fermi energy. Tunneling current is an 
integral of the LDOS bounded by the chemical potentials of the STM tip and 
surface electrodes, so it gives approximately the same values of the current 
for both voltage polarizations and therefore no topography reverse. In opposite 
situation, if the atomic energy is far away from the the Fermi energy, the LDOS 
is strongly asymmetric and centered around $\varepsilon_{\text{at}}$, thus 
giving a different contributions to the current. If $\varepsilon_{\text{at}}$ 
lays within $eU$ (between $\mu_{\text{STM}}$ and $\mu_{\text{surf}}$) tunneling 
occurs through the main peak of the LDOS giving large value of the current 
while in opposite situation the contribution is small as the tunneling takes 
place only through the tail of the main peak in the LDOS. 

\begin{figure}[h]
 \resizebox{0.8\linewidth}{!}{
  \includegraphics{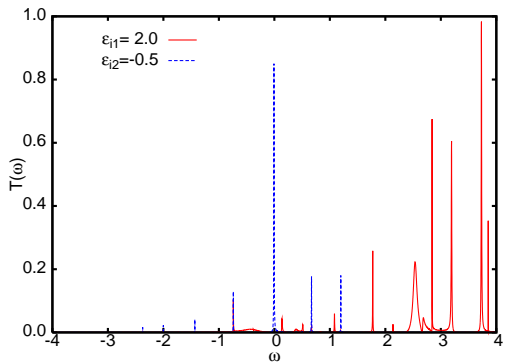}
  }
 \caption{\label{Fig9} (Color online) Energy dependence of the transmittance 
          ${\text{T}}(\omega)$ at STM tip position ${\text{x}} = 8$ in chain 
	  $1$ (solid line) and chain $2$ (dashed line). Model parameters are 
	  the same as in Fig. \ref{Fig8}.}
\end{figure}

Similar argument applies to the chain, however density of states (DOS) is now 
more complicated as it contains many peaks. Again, if the chain single particle 
energies lay close to the Fermi energy, the DOS is almost symmetric and thus no
topography reverse. If $\varepsilon_{i \alpha}$ are much larger than the Fermi
energy, density of states contains a series of peaks distributed mainly around
original values of $\varepsilon_{i \alpha}$, and therefore we get a large 
current for given bias sample and small one for reversed bias. This is clearly 
seen in Fig. \ref{Fig9}, where the example of the weighted DOS (transmittance) 
at STM tip position in the middle of the chain ${\text{x}} = 8$ (see Fig. 
\ref{Fig8}) is plotted as a function of energy. 

Turning to the experimental data, one can associate chain C with the chain $1$
in Fig. \ref{Fig8} (top panel) and chain D with that indicated by $2$ in Fig. 
\ref{Fig8} (bottom panel) and conclude that chain C (Fig. \ref{Fig2}) consists 
of single atoms with atomic energies much larger than the Fermi energy, while 
chain D has those energies close to the Fermi energy assuming that the other 
parameters characterizing chains are equal for both chains.

At this point it is interesting to note boundary effects clearly seen in Fig. 
\ref{Fig8} (bottom panel) and associated with the zero dimensional states at 
the ends of the chains, recently observed on Si(553)-Au surface \cite{Crain_2}.

\begin{figure}[h]
 \resizebox{0.9\linewidth}{!}{
  \includegraphics{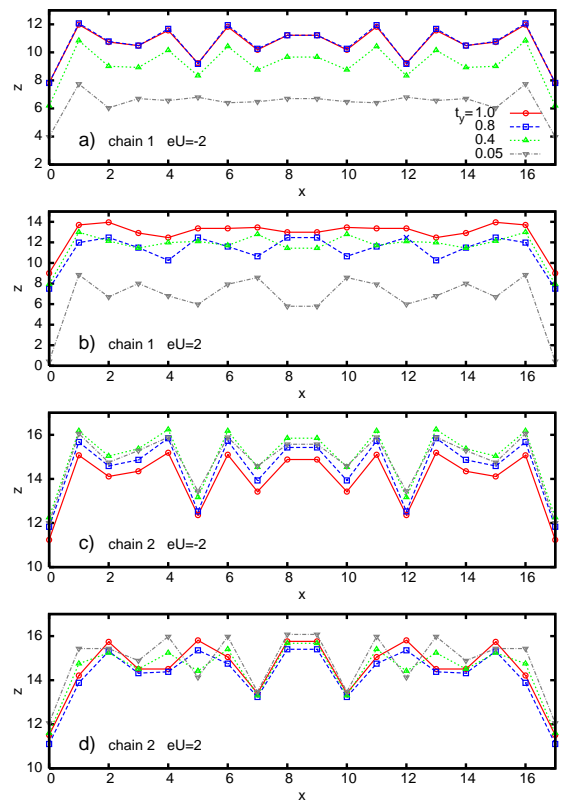}
  }
 \caption{\label{Fig10} (Color online) STM tip - surface distance along chain 
          $1$ for $eU = -2.0$ a) and $eU = 2.0$ b) and chain $2$ for 
	  $eU = -2.0$ c) and $eU = 2.0$ d) for different values of the 
	  inter-chain hopping integral $t_{\text{y}}$ indicated in the figure. 
	  Other parameters are $\varepsilon_0 = 0.0$, 
	  $t_{{\text{x}} 1} = t_{{\text{x}} 2} = 1.0$, 
	  $\varepsilon_{i 1} = 2.0$ and $\varepsilon_{i 2} = -0.5$.}
\end{figure}

Another important quantity is the inter-chain hopping integral $t_{\text{y}}$.
In Fig. \ref{Fig10} the STM tip - surface distance is plotted for different
values of the inter-chain couplings $t_{\text{y}}$.
It is clearly seen that the strength of the coupling can modify results. 
However, while it has strong influence on the properties of the chain $1$, 
eventually leading to the reverse of the topography for positive bias (compare 
curve with boxes ($t_{\text{y}} = 0.8$) with the others in the panel b)), it 
very weakly modifies topographic properties of the chain $2$. Such a difference 
in the influence can be explained in the following way. As we would expect the 
inter-chain hopping $t_{\text{y}}$ modifies the transmittance $T(\omega)$ of 
the system. It shifts the positions of the resonances in the $T(\omega)$ in 
both chains in the same way but at the same time it modifies their spectral 
weights in different way. However for $\varepsilon_{i \alpha}$ close to the
Fermi energy the transmittance integrated over the $eU$ remains almost
unchanged, thus leading to similar values of the tip - surface distance for 
different inter-chain couplings $t_{\text{y}}$.

\begin{figure}[h]
 \resizebox{0.9\linewidth}{!}{
  \includegraphics{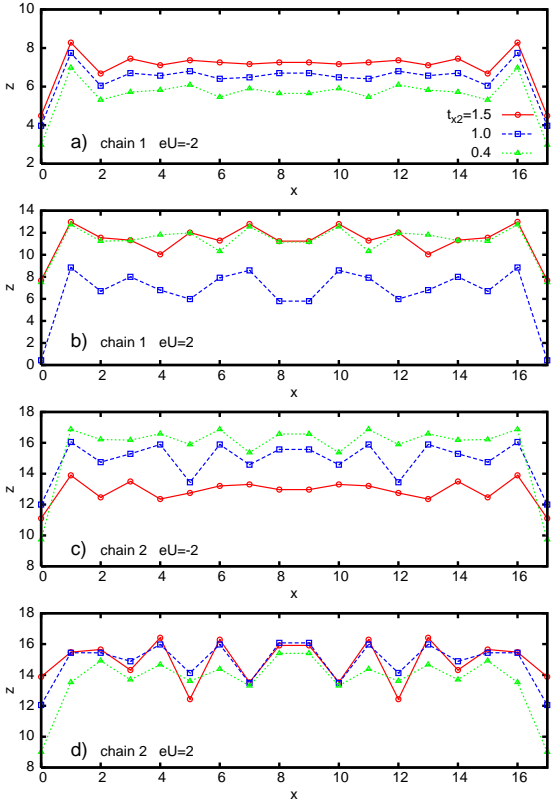}
  }
 \caption{\label{Fig11} (Color online) STM tip - surface distance along chain 
          $1$ for $eU = -2.0$ a) and $eU = 2.0$ b) and chain $2$ for 
	  $eU = -2.0$ c) and $eU = 2.0$ d) for different values of the hopping 
	  integral $t_{{\text{x}} 2}$ indicated in the figure and fixed 
	  $t_{{\text{x}} 1} = 1.0$. Other parameters are $\varepsilon_0 = 0.0$, 
	  $t_{\text{y}} = 0.05$, $\varepsilon_{i 1}=2.0$ and 
	  $\varepsilon_{i 2}=-0.5$.}
\end{figure}

Let us now discuss what the effect of the hopping integral between between
neighboring atoms in the same chain on the topography is. Figure \ref{Fig11} 
shows the topography of the chains for a number of the intra chain hopping 
integrals $t_{{\text{x}}2}$ with fixed $t_{{\text{x}}1} = 1.0$ and 
$\varepsilon_{i 1}=2.0$, $\varepsilon_{i 2}=-0.5$.
It is clearly seen that the intra chain hopping can change the topography of 
the chain. While for larger values of the hopping along chain $2$ than that 
along chain $1$ ($t_{{\text{x}} 2} > t_{{\text{x}} 1}$) one can observe reverse 
of the topography in both chains (compare curves with circles in all panels), 
for equal hoppings ($t_{{\text{x}} 2} = t_{{\text{x}} 1}$) there is a reverse 
of the topography in chain $1$ and no such effect in chain $2$ (curves with 
boxes) and for ($t_{{\text{x}} 2} < t_{{\text{x}} 1}$) there is no reverse of 
the topography in either chain (curves with triangles). Such modifications of 
the chain topography due to the intra chain hopping integrals is not a general 
property of the system as there is no similar effect in the symmetric 
situation, namely if the chain atomic energies were equal. We have performed a 
number of calculations for $\varepsilon_{i 1} = \varepsilon_{i 2}$ and got 
quite different behavior from that presented in Fig. \ref{Fig11}. Namely, our 
self-consistent calculations always converged to the results showing the 
reverse of the topography in both chains. This is a very important result as it 
leads to the conclusion that the chain atomic energies have to be different in 
both chains in order to get an agreement with the experimental data.

Similar effect on the topography has the coupling to the surface 
$V_{{\bf k} i \alpha}$. For $\varepsilon_{i 1} = \varepsilon_{i 2}$ one always
gets reverse of the topography in both chains. This can be easily understood if 
we remind that coupling to the surface leads to the broadening of the 
resonances in the density of states but does not influence their positions when 
the Coulomb interactions in chains are neglected. 

From above discussion one can conclude that the main effect which leads to the
reverse of the topography in one chain and its lack in the other one comes from
the chain atomic energies - they have to be different in both chains. Going
further one can conclude that the chains observed in STM have been made of 
different materials, one of silicon and the other one of gold. 


\section{\label{comparison} Comparison with experiment}

To make quantitative comparison of the theoretical calculations with the 
experimental data we use the formula for the tip - surface distance
(\ref{Gamma_z}) and put $\Gamma_{\text{surf}} = 1$ eV and 
$a = a_{[1 \bar{1} 1]} = 0.384$ nm. 

\begin{figure}[h]
 \resizebox{0.9\linewidth}{!}{
  \includegraphics{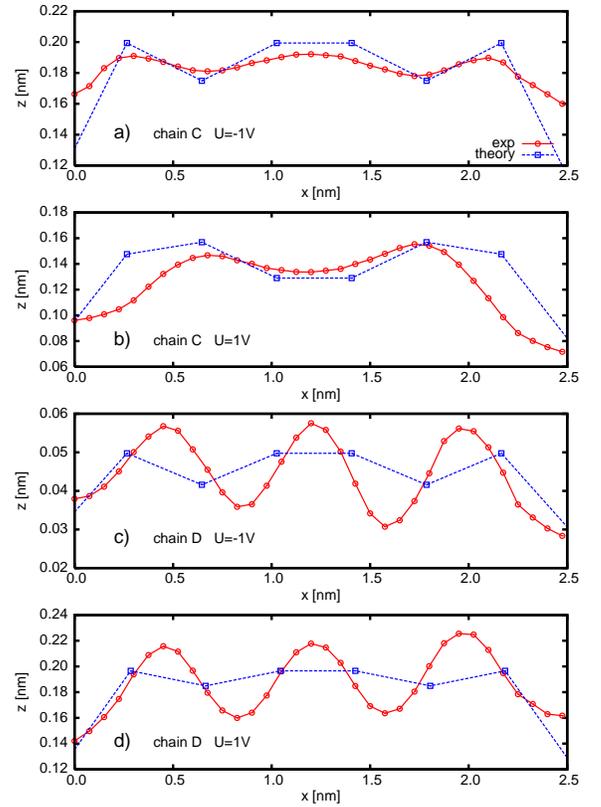}
  }
 \caption{\label{Fig12} (Color online) Comparison of the experimental data 
          (circles) of typical short chain C for $U = -1$ V a), $U = 1$ V b) 
	  and chain D for $U = -1$ V c) and $U = 1$ V d) with the theoretical 
	  calculations (squares). The model parameters are described in the 
	  text.}
\end{figure}

Figure \ref{Fig12} shows both the experimental and the theoretical data of 
typical short chains of type C and D, marked with smaller box in Fig. 
\ref{Fig1}. 
The best fit has been obtained with the chain atomic energies 
$\varepsilon_{i1} = 0.3$ eV for chain C and $\varepsilon_{i2} = - 0.2$ eV for 
chain D, respectively. Moreover, the values of the intra-chain and inter-chain 
hoppings were the same as those obtained in Ref. \cite{Crain_1}, i.e. 
$t_{{\text{x}} 1} = t_{{\text{x}} 2} = 0.68$ eV and $t_{{\text{y}}} = 0.05$ eV. 
Those values of the chain atomic energies, as it was stated previously, suggest 
that we deal with different chains, i.e. chains composed of different 
materials. Going further, we can assign the C chain to the Si atoms, while D
chain to the Au atoms. This can be deduced from different values of the atomic
energies $\varepsilon_{\alpha}$ used in calculations and electronegativity of 
free atoms. The Au atoms are more electronegative, so they should have lower 
atomic energies, $\varepsilon_{\text{at}} = \varepsilon_{i2} = -0.2$ eV in our 
case, while Si atoms have 
$\varepsilon_{\text{at}} = \varepsilon_{i1} = 0.3$ eV. Of course, this 
conclusion is true provided the other parameters are the same or similar in 
both chains. 

Although theoretical topography curves do not exactly follow experimental ones, 
main features, i.e. the reverse of the topography for chain C and its lack for 
chain D, are quite reasonably reproduced. The deviations between theoretical 
calculations and experimental data steam from the fact that we used a simple 
model of STM tunneling, nevertheless giving a semi-quantitative agreement with 
the experiment. Moreover, our model allows for calculations of the tunneling
current in the cases when the STM tip is placed exactly above the chain atom,
therefore we do not present the theoretical data for the tip position between
chain atoms. To get full agreement with experimental data one has to take into
account additional effects, like tunneling into aside chain atoms, the Coulomb 
interactions or perform first principles calculations which is out of the scope 
of the present work.


\section{\label{conclusions} Conclusions}

In conclusion we have studied properties of the mono-atomic chains on the 
vicinal Si(557)-Au surface using STM technique. We have observed double
non-equivalent chains on such surface showing different behavior of the
topography with reversing of the bias voltage. While chain C shows reverse of
the topography, chain D does not show such effect. Additionally the STM tip - 
surface distance oscillates along any chain with the same period equal to 
double lattice constant in direction $[1 \bar{1} 0]$. We have also performed
tight binding model calculations showing similar oscillations of the STM tip -
surface distance and led us to the conclusion that atomic energies have to be
different in different chains and larger in chain C than in chain D. This fact 
supports the scenario which is probably realized in experiment that one chain 
is made of gold (chain D) and the other one of silicon atoms (chain C).


\begin{acknowledgments}
This work has been supported by the KBN grant no. 1 P03B 004 28.
T. K. thanks the Foundation for Polish Science for a Fellowship for Young 
Scientists.
\end{acknowledgments}


\end{document}